\documentclass[a4paper]{article}
\title{Modified $N=1$ Green-Schwarz superstring with irreducible 
first class constraints}
\author{A.A. Deriglazov\thanks{alexei@fisica.ufjf.br ~ On leave of
absence from Dept. Math. Phys., Tomsk Polytechnical University,
Tomsk, Russia}}
\date{Instituto de F\'\i sica, Universidade Federal de Juiz de Fora,\\
MG, Brasil.}
\begin{document}
\maketitle
\large
\begin{abstract}
We propose modified action which is equivalent to $N=1$ Green-Schwarz 
superstring and which allows one to realize the supplementation trick 
[26]. Fermionic first and second class constraints are covariantly 
separated, the first class constraints (1CC) turn out to be irreducible. 
We discuss also equations of motion in the covariant gauge for 
$\kappa$-symmetry. It is shown how the usual Fock space picture can be 
obtained in this gauge.
\end{abstract}

{\bf PAC codes:} 0460D, 1130C, 1125 \\
{\bf Keywords:} Covariant quantization, Superstring \\

\noindent
\section{Introduction}
Manifestly super Poincare invariant formulation of branes implies 
appearance of mixed first and second class fermionic constraints 
in the Hamiltonian formalism (equivalently one has infinitely reducible 
local $\kappa$-symmetry in the Lagrangian formalism) [1-10]. Typically, 
first and second class constraints (2CC) are treated in a rather 
different way in quantum theory\footnote{Quantization scheme for mixed 
constraints was developed in [11]. Application of this scheme to concrete 
models may conflict with manifest Poincare covariance [12].}. In 
particular, to construct formal expression for the covariant path 
integral one needs to have splitted and irreducible constraints [13, 14]. 
So, it is necessary at first to split the constraints, 
which can be achieved 
by using of covariant projectors of one or other kind [15-17, 12]. 
Details depend on the model under consideration. For example, for 
CBS superparticle [18, 1] one introduces two auxilliary vector 
variables in addition to the initial superspace coordinates [15]. 
For the Green-Schwarz (GS) superstring the projectors can be constructed 
in terms of the initial variables only [16]. After that the problem 
reduces to quantization of the covariantly separated but 
{\em infinitely} reducible constraints. Despite a lot of efforts 
(see [15-23] and references therein) this problem has no fully 
satisfactory solution up to date. Namely, infinitely reducible 1CC 
imply infinite tower of ``ghost for ghosts'' variables [22]. For 
reducible 2CC the problem is (besides the problem of quantum 
realization) that the covariant Dirac bracket obeys the Jacobi identity 
on the second class constraints surface only [12]. A revival of interest 
to the problem is due to recent work [24] where it was shown that 
scattering amplitudes for superstring can be constructed in a manifestly 
covariant form, as well as due to progress in the light-cone 
quantization of superstring on $AdS_5\times S^5$ background [25, 24].

One possibility to avoid the problem of quantization of infinitely 
reducible constraints is the supplementation trick which was formulated 
in the Hamiltonian framework in [26]. The basic idea is to introduce 
additional fermionic variables subject to their own reducible 
constraints (the constraints are chosen in such a way that the additional 
sector do not contains physical degrees of freedom). Then the original 
constraints can be combined with one from the additional sector into 
irreducible set. For the latter one imposes a covariant and irreducible 
gauge. It implies, in particular, a possibility to construct correct 
Dirac bracket for the theory. 

Next problem arising in this context is the problem of linearisation 
of the physical sector dynamics. Crucial property of the standard 
noncovariant gauge $\Gamma^+\theta=0$ is that equations of motion in 
this case acquire linear form. Then it is possible to find their 
general solution. While not necessary for construction of the formal 
path integral, namely this fact allows one to fulfill really the 
canonical quantization procedure. Similarly to this, the covariant 
gauge will be reasonable only if it has the same property. 
Unfortunately, for the superstring in the covariant gauge for 
$\kappa$-symmetry and in the usual gauge for world sheet symmetry, 
equations of motion remain nonlinear. It will be shown that the 
problem can be avoided if one imposes ``off-diagonal gauge'' for 
$d=2$ fields. 

In symmary, the necessary steps toward to manifestly covariant 
formulation look as follows: mixed constraints $\mapsto$ separated but 
reducible constraints $\mapsto$ separeted irreducible constraints 
$\mapsto$ covariant irreducible gauge $\mapsto$ free dynamics of the 
physical sector. The final formulation admits application of the 
standard quantization methods in the covariant form.

To apply the recipe for concrete model one needs to find a modified 
Lagrangian action which reproduces the desired irreducible constraints. 
Some examples were considered [26, 27], in particular, the modified 
$N=1$ GS superstring  action was proposed in [28]. But it was pointed 
in [29]  that the action is not equivalent to the initial one (to split 
the constraints, an additional bosonic variables were introduced. Zero 
modes of the variables survive in the physical sector).

In this work we present modified action which is equivalent to $N=1$ 
GS superstring and which allows one to realize the supplementation 
trick. We consider $N=1$ theory as a toy example for type II GS 
superstring and restrict our attention to 1CC only. The reason is that, 
due to special structure of type II theory, its 2CC do not represent 
a problem and can be combined into irreducible set.

The work is organized as follows. To fix our notations, we review main 
steps of the supplementation scheme in Sec. 2 (see also [26]). 
In Sec. 3 we consider bosonic string (in ADM representation for $d=2$ 
metric) and discuss its dynamics in the off-diagonal gauge. It will be 
shown that the resulting Fock space picture of state spectrum is the 
same as in the standard gauge. We present also relation among string 
coordinates in these two gauges. The trick turns out to be necessary 
for linearisation of the superstring equations of motion in the 
covariant gauge for $\kappa$-symmetry. In Sec. 4 modified formulation 
of $N=1$ GS superstring action is presented and proved to be 
equivalent to the initial one. First and second class constraints are 
covariantly separated, 1CC form irreducible set. Equations of motion 
and their solution in the covariant gauge for $\kappa$-symmetry are 
discussed in Sec. 5. Some technical details are arranged in the 
Appendixes.  

\section{Supplementation of the reducible constraints.} 

It will be convenient to work in 16-component formalism of the Lorentz
group $SO(1, 9)$,
then $\theta^\alpha$, $\psi_\alpha$, $\alpha=1,\dots,16$, are
Majorana--Weyl spinors of opposite chirality. Real, symmetric
$16\times16$ $\Gamma$-matrices ${\Gamma^\mu}_{\alpha\beta}$,
$\tilde\Gamma^{\mu\alpha\beta}$ obey the algebra
$\Gamma^\mu\tilde\Gamma^\nu+\Gamma^\nu\tilde\Gamma^\mu=-2\eta^{\mu\nu},
~ \eta^{\mu\nu}=(+,-, \ldots ,-)$.
Momenta conjugate to configuration space variables $q^i$
are denoted as $p_{qi}$.

Let us consider a dynamical system with fermionic pairs $(\theta^\alpha,
p_{\theta\alpha})$ being presented among the phase space variables $z^A$.
Typical situation for the models under consideration is that the following
constraints
\begin{eqnarray}\label{1}
L_\alpha\equiv p_{\theta\alpha}- iB_\mu{\Gamma^\mu}_{\alpha\beta}
\theta^\beta\approx 0, \qquad D^\mu D_\mu\approx0,
\end{eqnarray}
are presented among others.
Here, the $B^\mu(z)$ and $D^\mu(z)$ are some functions of phase variables
$z$, so that $D^2\approx0$ is first class constraint.
Poisson bracket of the fermionic constraints is
\begin{eqnarray}\label{2}
\{L_\alpha,L_\beta\}=2iD_\mu{\Gamma^\mu}_{\alpha\beta}.
\end{eqnarray}
The system $L_\alpha\approx0$ is mixture of first and second class
constraints, as it will be proved momentarily. Supplementation scheme
for the mixed constraints consist of the following steps.

{\em A). Manifestly covariant separation of the constraints.} \\
Let us extend the initial phase space by
a pair of vectors $(\Lambda^\mu, p_{\Lambda\mu})$ subject to
constraints
\begin{eqnarray}\label{3}
\Lambda^2\approx0, \qquad p^\mu_\Lambda\approx0.
\end{eqnarray}
Supposing that $\Lambda D\ne0$ (which will be true for the models under
consideration), one can extract two 2CC:
$\Lambda^2\approx0$, $p_\Lambda D\approx0$ and nine 1CC: $\tilde
p^\mu_\Lambda\equiv p^\mu_\Lambda-\frac{p_\Lambda D}{\Lambda
D}\Lambda^\mu\approx0$ (there is identity $D\tilde
p_\Lambda\equiv0$). Thus the variables introduced are non physical. 
Eq.(\ref{3}) has first stage of reducibility 
and can be quantized by standard methods [34, 35].

Below the following two facts will be used systematically (proof is 
presented in the Appendix 2). \\
1). ~ Let $\Psi^\alpha=0$ are 16 equations. Then \\
a). ~ The system 
\begin{eqnarray}\label{4}
D^\mu\Gamma^\mu\Psi=0,
\end{eqnarray}
\begin{eqnarray}\label{5}
\Lambda^\mu\Gamma^\mu\Psi=0,
\end{eqnarray}
is equivalent to $\Psi^\alpha=0$. \\
b) ~ The quantities (\ref{4}), (\ref{5}) belong to $P_-, ~ P_+$ 
subspaces correspondingly. \\
c) ~ Let $\Psi^\alpha=0$ represent 16 independent equations. Then 
Eq.(\ref{4}) contains 8 independent equations. In $SO(8)$ notations 
they mean that $8_s$ part $\bar\Psi_{\dot a}$ of $\Psi^\alpha$ can be 
presented through $8_c$ part $\Psi_a$ (or vice-versa). The same is 
true for Eq.(\ref{5}). \\
2). ~ Let $\Psi^\alpha=0, ~ \Phi^\alpha=0$ are $16+16$ independent 
equations. Consider the system 
\begin{eqnarray}\label{6}
D^\mu\Gamma^\mu\Psi=0, \qquad
\Lambda^\mu\Gamma^\mu\Phi=0.
\end{eqnarray}
Then ~ a) Eq.(\ref{6}) contains 16 independent equations according 
to 1). \\
b) ~ The equations 
\begin{eqnarray}\label{7}
D^\mu\Gamma^\mu\Psi+\Lambda^\mu\Gamma^\mu\Phi=0,
\end{eqnarray}
are equivalent to the system (\ref{6}). Thus the system (\ref{7}) 
consist of 16 independent equations (i.e. it is irreducible). 

The system (\ref{1}) can be rewritten now in the equivalent form
\begin{eqnarray}\label{8}
L^{(1)\alpha}\equiv D_\mu\tilde\Gamma^{\mu\alpha\beta}L_\beta
\approx 0,
\end{eqnarray}
\begin{eqnarray}\label{9}
L^{(2)\alpha}\equiv \Lambda_\mu\tilde\Gamma^{\mu\alpha\beta}L_\beta
\approx 0.
\end{eqnarray}
The equivalence follows from the statement 1), or directly from 
invertibility of the matrix\footnote{To split the constraints one can 
also use true projectors instead of the matrices $D_\mu\Gamma^\mu, ~ 
\Lambda_\mu\Gamma^\mu$, see Appendix 2. It allows one to avoid possible 
``second class patalogy'' [30] in the 1CC algebra [16].}
$(\Lambda_\mu+D_\mu){\Gamma^\mu}_{\alpha\beta}$: ~
$(\Lambda_\mu+D_\mu){\Gamma^\mu}(L^{(1)}+L^{(2)})\approx -2(D\Lambda)L.$
Among 1CC $L^{(1)}\approx0$ and 2CC $L^{(2)}\approx0$ there are in
eight linearly independent (see Appendix 1).

{\em B). Auxiliary sector subject to reducible constraints.} \\ 
Let us further introduce a pair of spinors
$(\eta^\alpha,p_{\eta\alpha})$ subject to the constraints
\begin{eqnarray}\label{10}
p_{\eta\alpha}\approx0, \qquad T_\alpha\equiv \Lambda_\mu
{\Gamma^\mu}_{\alpha\beta}\eta^\beta\approx0.
\end{eqnarray}
These equations contain 8 independent 1CC among $p^{(1)}_\eta\equiv
\Lambda_\mu\tilde\Gamma^\mu p_\eta\approx0$ and 8+8 independent 2CC
among $p^{(2)}_\eta\equiv D_\mu\tilde\Gamma^\mu p_\eta\approx0$,
$\Lambda_\mu\Gamma^\mu\eta\approx0$. Note that the covariant gauge
$D_\mu\Gamma^\mu\eta=0$ may be imposed. After that the complete system
(constraints + gauges) is equivalent to $p_\eta\approx0$,
$\eta\approx0$.

{\em C). Supplementation up to irreducible constraints.} \\ 
Now part of the constraints can be combined into covariantly separated
and irreducible sets. According to the statement 2), the system 
(\ref{8})-(\ref{10}) is equivalent to
\begin{eqnarray}\label{11}
\Phi^{(1)\alpha)}\equiv L^{(1)\alpha}+p^{(1)\alpha}_\eta=D_\mu
\tilde\Gamma^{\mu\alpha\beta}L_\beta+\Lambda_\mu
\tilde\Gamma^{\mu\alpha\beta} p_{\eta\beta}\approx 0,
\end{eqnarray} 
\begin{eqnarray}\label{12}
\Phi^{(2)\alpha}\equiv L^{(2)\alpha}+p^{(2)\alpha}_\eta=\Lambda_\mu
\tilde\Gamma^{\mu\alpha\beta}L_\beta+D_\mu
\tilde\Gamma^{\mu\alpha\beta} p_{\eta\beta}\approx0, 
\end{eqnarray}
\begin{eqnarray}\label{13}
T_\alpha\equiv \Lambda_\mu{\Gamma^\mu}_{\alpha\beta}\eta^\beta
\approx0,
\end{eqnarray}
where $\Phi^{(1)\alpha}\approx0$ 
($\Phi^{(2)\alpha}\approx0$) are 16 irreducible 1CC
(2CC) and $T_\alpha\approx0$ include 8 linearly independent 2CC.
In the result first class constraints of the extended formulation form 
irreducible set (\ref{11}). As it was mentioned above the type II 
superstring presents an example of more attractive
situation as compare to the general case (\ref{11})-(\ref{13}). Due to
special structure of the theory the 2CC can also be combined into
irreducible set.

\section{Bosonic string in ``off-diagonal gauge''.}

In this section, on example of the bosonic string, we discuss two 
additional tools which will be used below. First, the modified 
superstring action acquires more elegant form in ADM representation 
for $d=2$ metric. Second, to analyze equations of motion in the covariant 
gauge for $\kappa$-symmetry it will be convenient to use a trick which 
we refer here as off-diagonal gauge for $d=2$ fields.

Starting from the bosonic string 
\begin{eqnarray}\label{14}
S=-\frac{T}{2}\int d^2\sigma
\frac{1}{\sqrt{-g}}g^{ab}
\partial_ax^\mu \partial_bx^\mu,
\end{eqnarray}
let us consider ADM representation
\begin{eqnarray}\label{15}
g^{00}=\frac{1}{\gamma N^2}, \quad g^{01}=\frac{N_1}{\gamma N^2}, \quad 
g^{11}=\frac{N_1^2-N^2}{\gamma N^2}, \quad 
\sqrt{-g}=\frac{1}{\gamma N}, 
\end{eqnarray}
then
\begin{eqnarray}\label{16}
N=\frac{\sqrt{-\det{g^{ab}}}}{g^{00}}, \qquad 
N_1=\frac{g^{01}}{g^{00}}, \qquad 
\gamma=\frac{-g^{00}}{\det{g^{ab}}}.
\end{eqnarray}
The action (\ref{14}) acquires now the form
\begin{eqnarray}\label{17}
S=-\frac{T}{2}\int d^2\sigma\frac{1}{N}
D_+x^\mu D_-x^\mu, \cr
D_{\pm}x^\mu\equiv\partial_0x^\mu+N_{\pm}\partial_1x^\mu, \qquad 
N_{\pm}\equiv N_1\pm N, 
\end{eqnarray}
while the world-sheet reparametrisations in this representation look as 
\begin{eqnarray}\label{18}
\delta\sigma^a=\xi^a, \qquad \delta N_{\pm}=\partial_0\xi^1+
(\partial_1\xi^1-
\partial_0\xi^0)N_{\pm}-\partial_1\xi^0N_{\pm}^2.
\end{eqnarray}
By direct application of the Dirac procedure one obtains the Hamiltonian 
\begin{eqnarray}\label{19}
H=\int d\sigma\left[-\frac N2\left(\frac {1}{T} p^2+
T(\partial_1 x)^2\right)-N_1(p\partial_1 x)+
\lambda_N p_N+\lambda_{N1}p_{N1}\right],
\end{eqnarray}
where $\lambda_q$ are the Lagrangian multipliers for the corresponding 
primary constraints. Dynamics is governed by the equations of motion
\begin{eqnarray}\label{20}
\partial_0x^\mu=-\frac{N}{T}p^\mu-N_1\partial_1x^\mu, \quad 
\partial_0p^\mu=-\partial_1[TN\partial_1x^\mu+N_1p^\mu],
\end{eqnarray}
which are accompanied by the first class constraints
\begin{eqnarray}\label{21}
p_N=0, \qquad p_{N1}=0,
\end{eqnarray}
\begin{eqnarray}\label{22}
(\frac{1}{T}p^\mu\pm\partial_1x^\mu)^2=0.
\end{eqnarray}
In the standard gauge 
\begin{eqnarray}\label{23}
N=1, \qquad N_1=0,
\end{eqnarray}
one has 
\begin{eqnarray}\label{24}
\partial_0x^\mu=-\frac{1}{T}p^\mu, \qquad
\partial_0p^\mu=-T\partial_1\partial_1x^\mu,
\end{eqnarray}
which implies $(\partial_0^2-\partial_1^2)x^\mu=0, ~ 
(\partial_0^2-\partial_1^2)p^\mu=0$. Now one can look for solution in 
terms of oscillators. Let us show that, instead of this, one can 
equivalently to start from the system $\partial_-\tilde x^\mu=0, ~ 
\partial_-\tilde p^\mu=0$ and look for its general 
solution\footnote{It follows immediately from the wave equation for 
$x^\mu$. We prefer to work with the Hamiltonian equations of motion 
since it gives automatically brackets for oscillators.}. From the 
latter one restores the general solution of Eq.(\ref{24}). Actually, 
Eq.(\ref{24}) can be rewritten as 
\begin{eqnarray}\label{25}
\partial_+x^\mu=-\frac{1}{T}\tilde p^\mu,
\end{eqnarray}
\begin{eqnarray}\label{26}
\partial_-\tilde p^\mu=0,
\end{eqnarray} 
where $\tilde p^\mu\equiv p^\mu-T\partial_1x^\mu$. Eq.(\ref{26}) has 
the solution $\tilde p^\mu(\tau, \sigma)=\tilde p^\mu(\sigma^+)$. 
Then (\ref{25}) is ordinary differential equation with the solution 
$x^\mu=z^\mu-\frac{1}{T}\int_{0}^{\sigma^+}dl\tilde p^\mu(l)$, where 
$z^\mu$ is general solution for $\partial_+z^\mu=0$. Equivalently, one 
solves $\partial_-\tilde x^\mu=0$, then $z^\mu=\tilde x^\mu
(\sigma^+\mapsto\sigma^-)$. Collecting all this one has the following 
result: \\
Let $\tilde x^\mu, ~ \tilde p^\mu$ represent general solution of 
the system
\begin{eqnarray}\label{27}
\partial_-\tilde x^\mu=0, \qquad 
\partial_-\tilde p^\mu=0. 
\end{eqnarray}
Then the quantities
\begin{eqnarray}\label{28}
x^\mu(\tau, \sigma)=\tilde x^\mu(\sigma^+\mapsto\sigma^-)-
\frac{1}{T}\int_{0}^{\sigma^+}dl\tilde p^\mu(l), \cr 
p^\mu(\tau, \sigma)=\frac 12[\tilde p^\mu-T\partial_-\tilde x^\mu
(\sigma^+\mapsto\sigma^-)],
\end{eqnarray}
give general solution of the system (\ref{24}).

Note that Eq.(\ref{27}) can be obtained from Eq.(\ref{20}) if one 
takes\footnote{Curious fact is that the membrane equations of motion 
turn out to be free also for similar choice.} 
\begin{eqnarray}\label{29}
N=0, \qquad N_1=-1,
\end{eqnarray}
instead of the gauge (\ref{23}). Note also that while the action 
(\ref{17}) is not well defined for the value $N=0$, the Hamiltonian 
formulation (\ref{19})-(\ref{22}) admits formally Eq.(\ref{29}) as 
the gauge fixing conditions for the constraints (\ref{21}).

One more observation is that the transition (\ref{28}) to the initial 
variables is not necessary in the canonical quantization framework. 
Namely, solution of Eq.(\ref{27}) in terms of oscillators leads to the 
same description of state space as those of (\ref{24}). Actually, 
solution of Eq.(\ref{27}) is ($0\le\sigma\le\pi$, closed string)
\begin{eqnarray}\label{30}
x^\mu(\tau, \sigma)=X^\mu+\frac{i}{\sqrt{\pi T}}\sum_{n\ne 0}\frac 1n
\beta_n^\mu e^{2in(\tau+\sigma)}, \cr
p^\mu(\tau, \sigma)=
\frac{1}{\pi}P^\mu-2\sqrt{\frac{T}{\pi}}\sum_{n\ne 0}
\gamma_n^\mu e^{2in(\tau+\sigma)}.
\end{eqnarray}
From these expressions one extracts the Poisson brackets for  
coefficients. For the variables 
\begin{eqnarray}\label{31}
\bar\alpha_n^\mu\equiv\beta_n^\mu+\gamma_n^\mu, \quad 
\alpha_n^\mu\equiv\beta_{-n}^\mu-\gamma_{-n}^\mu, \quad
\alpha_0^\mu=-\bar\alpha_0^\mu=\frac{1}{2\sqrt{\pi T}}P^\mu,
\end{eqnarray}
one obtains the properties
\begin{eqnarray}\label{32}
\{\alpha_n^\mu, \alpha_k^\nu\}=
\{\bar\alpha_n^\mu, \bar\alpha_k^\nu\}=
in\eta^{\mu\nu}\delta_{n+k,0}, \quad 
\{X^\mu, P^\nu\}=\eta^{\mu\nu}, \cr
(\alpha_n^\mu)^*= \alpha_{-n}^\mu, \qquad 
(\bar\alpha_n^\mu)^*=\bar \alpha_{-n}^\mu. 
\end{eqnarray}
In terms of these variables the Virasoro constraints (\ref{22}) 
acquire the standard form
\begin{eqnarray}\label{33}
L_n=\frac 12\sum_{\forall k}\alpha_{n-k}\alpha_k=0, \quad
\bar L_n=\frac 12\sum_{\forall k}\bar\alpha_{n-k}
\bar\alpha_k.
\end{eqnarray}
Eqs.(\ref{32}), (\ref{33}) have the same form as those of string in the 
gauge (\ref{23}) [31-33]. Thus, instead of the standard gauge one 
can equivalently use the conditions (\ref{29}), which gives the same 
structure of state space. This fact will be used in Sec. 5 for the 
superstring. If it is necessary, the initial string coordinate $x^\mu$ 
can be restored by means of Eq.(\ref{28}).

\section{$N=1$ Green-Schwarz superstring with irreducible first 
class constraints.}

Consider GS superstring action with $N=1$ space time supersymmetry
\begin{eqnarray}\label{34}
S=-\frac{T}{2}\int d^2\sigma
\left[\frac{1}{\sqrt{-g}}g^{ab}
\Pi_a^\mu \Pi_b^\mu+2i\varepsilon^{ab}\partial_ax^\mu
\theta\Gamma^\mu\partial_b\theta\right],
\end{eqnarray}
where $\sqrt{-g}=\sqrt{-\det{g^{ab}}}, ~ \Pi^\mu_a\equiv
\partial_ax^\mu-i\theta\Gamma^\mu\partial_a\theta, ~ 
\varepsilon^{01}=-1$. Let us denote
\begin{eqnarray}\label{35}
B^\mu\equiv p^\mu+T\Pi_1^\mu, \qquad 
\hat p^\mu\equiv p^\mu-iT\theta\Gamma^\mu\partial_1\theta, \cr 
D^\mu\equiv\hat p^\mu+T\Pi_1^\mu=
p^\mu+T\partial_1x^\mu-2iT\theta\Gamma^\mu\partial_1\theta, \cr 
\Lambda^\mu\equiv\hat p^\mu-T\Pi_1^\mu=p^\mu-T\partial_1x^\mu.
\end{eqnarray}
Then constraints under the interest for Eq.(\ref{34}) can be written as 
\begin{eqnarray}\label{36}
L_\alpha\equiv p_{\theta\alpha}- iB_\mu{\Gamma^\mu}_{\alpha\beta}
\theta^\beta\approx 0, \qquad D^2\approx0,
\end{eqnarray}
\begin{eqnarray}\label{37}
\Lambda^2\approx0.
\end{eqnarray}
From Eqs.(\ref{35})-(\ref{36}) and from the standard requirement that 
the induced metric is non degenerated it follows 
\begin{eqnarray}\label{38}
D\Lambda=\hat p^2-T\Pi_1^2\ne 0.
\end{eqnarray}
The constraints obey the algebra 
\begin{eqnarray}\label{39}
\{L_\alpha,L_\beta\}=2iD_\mu{\Gamma^\mu}_{\alpha\beta}
\delta(\sigma-\sigma ');
\end{eqnarray}
\begin{eqnarray}\label{40}
\{D^2, D^2\}=4T[D^2(\sigma)+D^2(\sigma ')]
\partial_\sigma\delta(\sigma-\sigma '), \cr 
\{\Lambda^2, \Lambda^2\}=-4T[\Lambda^2(\sigma)+\Lambda^2(\sigma ')]
\partial_\sigma\delta(\sigma-\sigma '), \cr
\{D^2, \Lambda^2\}=0;
\end{eqnarray}
\begin{eqnarray}\label{41}
\{L_\alpha, D^2\}=8iTD^\mu(\Gamma^\mu\partial_1\theta)_\alpha
\delta(\sigma-\sigma ')
\end{eqnarray}
From comparison of Eqs(\ref{36})-(\ref{39}) with Eqs.(\ref{1})-(\ref{3}) 
one concludes that the first step of supplementation scheme is 
{\em not necessary} here, since the quantity $\Lambda^\mu$ with the 
properties $\Lambda^2=0, ~ (D\Lambda)\ne 0$ is constructed from the 
variables in our disposal. Thus one needs to find only a modification 
which will lead to Eq.(\ref{10}). 

To achieve this it will be convenient to work in the ADM 
representation (\ref{15}). The action (\ref{34}) acquires then the 
following form:
\begin{eqnarray}\label{42}
S=-\frac{T}{2}\int d^2\sigma
\left[\frac{1}{N}
\Pi_+^\mu \Pi_-^\mu+2i\varepsilon^{ab}\partial_ax^\mu
\theta\Gamma^\mu\partial_b\theta\right],
\end{eqnarray}
where it was denoted 
\begin{eqnarray}\label{43}
\Pi^\mu_{\pm}\equiv\Pi_0^\mu+N_{\pm}\Pi_1^\mu, \qquad 
N_{\pm}\equiv N_1\pm N.
\end{eqnarray}
Modified action to be examined is
\begin{eqnarray}\label{44}
S=-\frac{T}{2}\int d^2\sigma
\left[\frac{1}{N}
\Pi_+^\mu (\Pi_-^\mu+i\eta\Gamma^\mu\chi)+
2i\varepsilon^{ab}\partial_ax^\mu
\theta\Gamma^\mu\partial_b\theta-\frac{1}{4N}
(\eta\Gamma^\mu\chi)^2\right],
\end{eqnarray}
where two additional Majorana-Weyl fermions 
$\eta^\alpha(\tau, \sigma), ~ \chi^\alpha(\tau, \sigma)$ were introduced. 
Our aim now will be to show canonical equivalence of this action and 
the initial one. Then the additional sector will be used to arrange 
1CC of the theory into irreducible set.

Direct application of the Dirac algorithm gives us the Hamiltonian
\begin{eqnarray}\label{45}
H=\int d\sigma\left[-\frac N2\left(\frac {1}{T} \hat p^2+
T\Pi_1^2\right)-N_1(\hat p\Pi_1)-
\frac i2 (\hat p^\mu-T\Pi_1^\mu)\eta\Gamma^\mu\chi+\right. \cr
\left.\lambda_N p_N+\lambda_{N1}p_{N1}+\lambda_\eta p_\eta+
\lambda_\chi p_\chi+L\lambda_\theta\right],
\end{eqnarray}
where $\lambda_q$ are the Lagrangian multipliers for the corresponding 
primary constraints. After determining of secondary constraints, complete 
constraint system can be written in the following form (in the notations 
(\ref{35}))
\begin{eqnarray}\label{46}
p_N=0, \qquad p_{N1}=0,
\end{eqnarray}
\begin{eqnarray}\label{47}
H_+\equiv D^2-4TL\partial_1\theta=0, \quad 
H_-\equiv \Lambda^2-4T\partial_1\eta p_\eta-4T\partial_1\chi p_\chi=0, 
\end{eqnarray}
\begin{eqnarray}\label{48}
L_\alpha\equiv p_{\theta\alpha}- iB_\mu{\Gamma^\mu}_{\alpha\beta}
\theta^\beta=0, 
\end{eqnarray}
\begin{eqnarray}\label{49}
G_\alpha\equiv\Lambda^\mu(\Gamma^\mu\eta)_\alpha=0, \qquad 
p_{\eta\alpha}=0.
\end{eqnarray}
\begin{eqnarray}\label{50}
S_\alpha\equiv\Lambda^\mu(\Gamma^\mu\chi)_\alpha=0, \qquad
p_{\chi\alpha}=0.
\end{eqnarray}
Note that the desired constraints (\ref{10}) appear in duplicate form 
(\ref{49}), (\ref{50}). Combinations of the constraints in 
Eq.(\ref{47}) are chosen in such a way that all mixed brackets 
(i.e. those among Eq.(\ref{47}) and Eqs.(\ref{48})-(\ref{50})) vanish. 
The constraints $H_{\pm}, ~ L_\alpha$ obey the Poisson bracket algebra 
(\ref{39}), (\ref{40}), while the remaining non zero brackets are 
written in the Appendix 3. The constraints (\ref{46}), (\ref{47}) are 
first class. Important moment is that there are no of tertiary 
constraints in the problem. Actually, from the condition that the 
constraints (\ref{48})-(\ref{50}) are conserved in time, one obtains 
\begin{eqnarray}\label{51}
D^\mu\Gamma^\mu(\lambda_\theta-N_+\partial_1\theta)=0, 
\end{eqnarray}
\begin{eqnarray}\label{52} 
Z^\alpha\equiv\Lambda^\mu(\Gamma^\mu\lambda_\eta)_\alpha+
iT[\partial_1(\eta\Gamma^\mu\chi)](\Gamma^\mu\eta)_\alpha=0,
\end{eqnarray}
and the same as (\ref{52}) for $\lambda_\chi$. Eq.(\ref{51}) allows 
one to determine half of the multipliers $\lambda_\theta$. According 
to statement 1), Eq.(\ref{52}) is equivalent to 
\begin{eqnarray}\label{53}
\Lambda^\mu\tilde\Gamma^\mu Z=0,
\end{eqnarray}
\begin{eqnarray}\label{54}
D^\mu\tilde\Gamma^\mu Z=0.
\end{eqnarray}
One finds that (\ref{53}) vanishes on the constraint surface, while 
manifest form of (\ref{54}) is 
\begin{eqnarray}\label{55}
2(D\Lambda)\tilde P^\alpha_{-\beta}\lambda^\beta_\eta=iTD^\mu
[\partial_1(\eta\Gamma^\nu\chi)](\tilde\Gamma^\mu\Gamma^\nu\eta)^\alpha.
\end{eqnarray}
Here $\tilde P_{\pm}$ are covariant projectors (A.12) on eight-dimensional 
subspaces. From Eqs.(\ref{54}), (\ref{55}), (A.17) it follows that 
both sides of Eq.(\ref{55}) belong to the same subspace $\tilde P_-$. 
So, Eq.(\ref{55}) do not contains of new constraints and allows one to 
determine half of the multipliers $\lambda_\eta$.

To proceed further, let us make partial fixation of gauge. One imposes 
$N=1, ~ N_1=0$ for Eq.(\ref{46}) and $D^\mu\Gamma^\mu\chi=0$ for 
1CC $\Lambda^\mu\Gamma^\mu p_\chi=0$ contained in Eq.(\ref{50}). After 
that, the pairs $(N, p_N), (N_1, p_{N1}), (\chi, p_\chi)$ can be omitted 
from consideration. The Dirac bracket for the remaining variables 
coincides with the Poisson one. In the same fashion, the pair 
$\eta, p_\eta$ can be omitted also. Then the remaining constraints (as 
well as equations of motion) coincide with those of the GS superstring, 
which proves equivalence of the actions (\ref{34}) and (\ref{44}).

On other hand, retaining the constraints (\ref{49}), the system 
(\ref{48}), (\ref{49}) can be rewritten equivalently as in 
(\ref{11})-(\ref{13}), or, in the manifest form 
\begin{eqnarray}\label{56}
\Phi^\alpha\equiv(\hat p^\mu+T\Pi_1^\mu)(\tilde\Gamma^\mu L)^\alpha+
(\hat p^\mu-T\Pi_1^\mu)(\tilde\Gamma^\mu p_\eta)^\alpha=0, 
\end{eqnarray}
\begin{eqnarray}\label{57}
(\hat p^\mu-T\Pi_1^\mu)(\tilde\Gamma^\mu L)^\alpha+
(\hat p^\mu+T\Pi_1^\mu)(\tilde\Gamma^\mu p_\eta)^\alpha=0, \cr
(\hat p^\mu-T\Pi_1^\mu)(\Gamma^\mu\eta)_\alpha=0,
\end{eqnarray}
with the irreducible 1CC (\ref{56}) which are separated from the
2CC (\ref{57}).

Covariant and irreducible gauge for Eq. (\ref{56}) can be chosen as 
\begin{eqnarray}\label{58}
R_\alpha\equiv\Lambda^\mu(\Gamma^\mu\theta)_\alpha+
D^\mu(\Gamma^\mu\eta)_\alpha=0,
\end{eqnarray}
or, equivalently
\begin{eqnarray}\label{59}
\Lambda^\mu(\Gamma^\mu\theta)_\alpha=0, \qquad 
D^\mu(\Gamma^\mu\eta)_\alpha=0.
\end{eqnarray}
Matrix of the Poisson brackets 
\begin{eqnarray}\label{60} 
\{\Phi^\alpha, R_\beta\}=[2(D\Lambda)\delta^\alpha_\beta+
4iTD^\mu(\tilde\Gamma^\mu\Gamma^\nu\partial_1\theta)^\alpha
(\Gamma^\nu\eta)_\beta]\delta(\sigma-\sigma '), 
\end{eqnarray}
has a body on its diagonal and is invertible. It means that 
Eqs.(\ref{56}), (\ref{58}) allows one to construct the Dirac bracket 
without fermionic patalogies.

\section{Superstring dynamics in the covariant gauge 
for $\kappa$-symmetry.}

To study equations of motion for physical variables let us consider 
the gauge $D^\mu\Gamma^\mu\chi=0$ and Eq.(\ref{59}). Then the variables 
$(\chi, p_\chi), (\eta, p_\eta)$ can be omitted. Equations of motion for 
the theory look now as follows
\begin{eqnarray}\label{61}
\partial_0x^\mu=-\frac{N}{T}p^\mu-N_1\partial_1x^\mu-
i\theta\Gamma^\mu(\lambda_\theta-N_+\partial_1\theta), \cr
\partial_0p^\mu=\partial_1[-TN\partial_1x^\mu-N_1p^\mu- 
iT\theta\Gamma^\mu(\lambda_\theta-N_+\partial_1\theta)], \cr
\partial_0\theta=-\lambda_\theta^\alpha.
\end{eqnarray}
The remaining constraints are (\ref{46})-(\ref{48}), which are 
acompanied by the covariant gauge condition 
\begin{eqnarray}\label{62}
\Lambda^\mu\Gamma^\mu\theta=0.
\end{eqnarray}
Its conservation in time gives the condition 
\begin{eqnarray}\label{63}
\Lambda^\mu\Gamma^\mu(\lambda_\theta-N_-\partial_1\theta)=0.
\end{eqnarray}
The Lagrangian multipliers $\lambda_\theta$ can be determined now from 
Eqs.(\ref{51}), (\ref{63}) 
\begin{eqnarray}\label{64}
\lambda_\theta^\alpha=N_1\partial_1\theta^\alpha+
N\tilde K^\alpha{}_\beta\partial_1\theta^\beta.
\end{eqnarray}
By using of this result in Eq.(\ref{61}) one has 
\begin{eqnarray}\label{65}
\partial_0x^\mu=-\frac{N}{T}p^\mu-N_1\partial_1x^\mu+
2iN\theta\Gamma^\mu\tilde P_-\partial_1\theta, \cr
\partial_0p^\mu=\partial_1[-TN\partial_1x^\mu-N_1p^\mu+
2iNT\theta\Gamma^\mu\tilde P_-\partial_1\theta], \cr
\partial_0\theta^\alpha=-N_1\partial_1\theta^\alpha-
\tilde K^\alpha{}_\beta\partial_1\theta^\beta.
\end{eqnarray}
In the standard gauge $N=1, ~ N_1=0$ for the constraints (\ref{46}) one 
obtains equations of motion in the following form: 
$\partial_0\theta^\alpha=-\tilde K^\alpha{}_\beta\partial_1\theta^\beta$, 
~ $\partial_0x^\mu=-\frac{1}{T}p^\mu+i\theta\Gamma^\mu\partial_+\theta,$ 
~ $\partial_0p^\mu=-T\partial_1[\partial_1x^\mu-
i\theta\Gamma^\mu\partial_+\theta]$. Thus, they remain nonlinear. 
Moreover, in this gauge we failed to find an appropriate set of 
variables that would obey to the free equations (the only quantity with 
desired property is $\Lambda^\mu$: ~ $\partial_-\Lambda^\mu=0$).

Nevertheless, usual picture of the Fock space can be obtained in the 
covariant gauge. To resolve the problem one can use the same trick which 
was considered above for the bosonic case. Namely, after substitution 
$N=0, ~ N_1=-1$ (equivalent choice is $N_1=1$) into Eq.(\ref{65}), one 
obtains free equations of motion
\begin{eqnarray}\label{66}
\partial_-x^\mu=0, \qquad \partial_-p^\mu=0, \qquad
\partial_-\theta_a=0, 
\end{eqnarray}
where $\theta_a, ~ a=1, \ldots ,8$ is $8_c$ part of $\theta^\alpha$, 
while $8_s$ part $\theta_{\dot a}$ is determined by the covariant gauge 
condition $\Lambda^\mu\Gamma^\mu\theta=0$. Fermionic dynamics is the same 
as in the light-cone gauge $\Gamma^+\theta=0$. Bosonic sector leads to 
correct description as it was proved in Sec. 3. Thus the covariant 
gauge (\ref{62}) allows one to obtain the same structure of state space 
as those in the light-cone gauge.   

\section{Conclusion}

In this work we have proposed modified action (\ref{44}) for $N=1$ GS 
superstring. In addition to usual superspace coordinates it involves a 
pair of the Majorana-Weyl spinors. The additional variables are subject 
to reducible constraints (\ref{49}), (\ref{50}) which supply their 
nonphysical character (see discussion after Eq.(\ref{10})). Equivalence 
of the modified action and the initial one was proved in the canonical 
quantization framework (see discussion after Eq.(\ref{55})). We have 
demonstrated also how the state spectrum can be studied in the covariant 
gauge for $\kappa$-symmetry.

In the modified formulation first class constraints form irreducible set 
(\ref{56}) and are separated from the second class one (\ref{57}). The 
corresponding covariant gauge (\ref{58}) is irreducible also, which 
garantees applicability of the usual path integral methods for the 1CC 
sector of the theory. In the considered theory with one supersymmetry, 
reducibility of the second class constraints can not be avoided. But it 
turns out to be possible for type IIB GS superstring. For this case the  
theory has two copies of the fermionic constraints (which correspond 
to two $\theta^A, ~ A=1,2$) with the same chirality. It allows one to 
consider their Poincare covariant combinations. In this case both first 
and second class constraints can be arranged into covariant sets in the 
initial formulation. For the type IIA theory the two copies of 
constraints have an opposite chirality and can not be combined in the 
initial formulation. Repeating the same procedure as in $N=1$ case, one 
finds that all the second class constraints  
can be combined into irreducible sets.   
These results will be presented in a forthcoming work.

\section*{Acknowledgments.}

The work has been supported partially by Project GRACENAS No 97-6.2-34.
 
\section*{Appendix 1. $SO(1, 9)$ and $SO(8)$ notations}
\setcounter{equation}{0}
\def\theequation{A.\arabic{equation}}
Manifest expression of $SO(1, 9)$, $\Gamma$-matrices $16\times16$ 
through $SO(8)$ $\gamma$-matrices is
\begin{eqnarray}
&& \Gamma^0=\left(\begin{array}{cc} {\bf 1}_8 & 0\\
0 & {\bf 1}_8\end{array}\right),\;\;
 \tilde\Gamma^0=\left(\begin{array}{cc} -{\bf 1}_8 & 0\\
0 & -{\bf 1}_8\end{array}\right),\cr
&& \Gamma^i=\left(\begin{array}{cc} 0 & {\gamma^i}_{a\dot a}\\
\bar\gamma^i{}_{\dot aa} & 0 \end{array}\right), \;\;
 \tilde\Gamma^i=\left(\begin{array}{cc} 0 & {\gamma^i}_{a\dot a}\\
\bar\gamma^i{}_{\dot aa} & 0\end{array}\right),\cr
&& \Gamma^9=\left(\begin{array}{cc} {\bf 1}_8 & 0\\
0 & -{\bf 1}_8\end{array}\right), \;\;
 \tilde\Gamma^9=\left(\begin{array}{cc} {\bf 1}_8 & 0\\
0 & -{\bf 1}_8\end{array}\right).
\end{eqnarray}
Here ${\gamma^i}_{a\dot a}$, $\bar\gamma^i{}_{\dot aa}\equiv
({\gamma^i}_{a\dot a})^{\rm T}$ are real $SO(8)$ $\gamma$-matrices
which obey [31]
\begin{equation}
\gamma^i\bar\gamma^j+\gamma^j\bar\gamma^i=2\delta^{ij}{\bf 1}_8,
\end{equation}
and $i,a,\dot a=1,\dots,8$. Majorana-Weyl spinors of $SO(1, 9)$ group 
$\Psi^\alpha, ~ \Phi_\alpha$ can be decomposed in terms of their 
$SO(8)$ components. Namely, from Eq.(A.1) it follows that in the 
decomposition 
\begin{eqnarray}
\Psi^\alpha=(\Psi_a, \bar\Psi_{\dot a}), \qquad 
\Phi^\alpha=(\Phi_a, \bar\Phi_{\dot a}),
\end{eqnarray}
the part $\Psi_a ~ (\bar\Psi_{\dot a})$ is $8_c ~ (8_s)$ 
representation of 
$SO(8)$ group. The matrices $\Gamma^{\pm}=\frac 12 (\Gamma^0\pm\Gamma^9)$ 
can be used to extract these components: $\Gamma^+\Psi\subset 8_c$, 
$\Gamma_-\Psi\subset 8_s$. It breaks $SO(1, 9)$ symmetry up to $SO(8)$ 
subgroup. To keep $SO(1, 9)$ group one needs to use covariant projectors 
described in Appendix 2. 

Fermionic constraints (\ref{1})  in $SO(8)$ notations are 
\begin{eqnarray}
L_a=p_{\theta a}-i(\sqrt 2B^-\theta_a-
B^i\gamma^i_{a\dot a}\bar\theta_{\dot a})=0, \cr 
\bar L_{\dot a}=\bar p_{\theta\dot a}-i(\sqrt 2B^+\bar\theta_{\dot a}-
B^i\bar\gamma^i_{\dot aa}\theta_a)=0,
\end{eqnarray}
and obey the algebra
\begin{eqnarray}
\{L_a, L_b\}=2\sqrt 2iD^-\delta_{ab}, \qquad 
\{\bar L_{\dot a}, \bar L_{\dot b}\}=
2\sqrt 2iD^+\delta_{\dot a\dot b}, \cr
\{L_a, \bar L_{\dot b}\}=-2iD^i\gamma^i_{a\dot b},
\end{eqnarray}
as a consequence of Eq.(\ref{2}). Thus they have no of definite class. 
In the separated form (\ref{8}), (\ref{9}) one has 
\begin{eqnarray}
\{L^{(1)\alpha}, L^{(1)\beta}\}=
\{L^{(1)\alpha}, L^{(2)\beta}\}\approx 0, \cr
\{L^{(2)\alpha}, L^{(2)\beta}\}\approx
-4i(D\Lambda)(\Lambda\tilde\Gamma)^{\alpha\beta},
\end{eqnarray}
on the constraint surface. Eqs.(\ref{8}), (\ref{9}) contains in eight 
linearly independent equations according to the statement 1). In 
$SO(8)$ notations one has
\begin{eqnarray}
L^{(1)}_a=-\sqrt 2D^+L_a-D^i\gamma^i_{a\dot a}\bar L_{\dot a}, \quad 
\bar L^{(1)}_{\dot a}=-\sqrt 2D^-\bar L_{\dot a}-
D^i\bar\gamma^i_{\dot aa}L_a, \cr 
L^{(2)}_a=-\sqrt 2\Lambda^+L_a-
\Lambda^i\gamma^i_{a\dot a}\bar L_{\dot a}, \quad
\bar L^{(2)}_{\dot a}=-\sqrt 2\Lambda^-\bar L_{\dot a}-
\Lambda^i\bar\gamma^i_{\dot aa}L_a.
\end{eqnarray}
One can take $L^{(1)}_a=0, ~ L^{(2})_a=0$ as linearly independent sets of 
first and second class constraints. Then the corresponding non zero 
bracket
\begin{eqnarray}
\{L^{(2)}_a, L^{(2)}_b\}\approx 
-4\sqrt 2i(D\Lambda)\Lambda^-\delta_{ab},
\end{eqnarray}
is manifestly invertible.

\section*{Appendix 2. Covariant projectors and their properties.}
\setcounter{equation}{8}
\def\theequation{A.\arabic{equation}}
To extract $8_c$ or $8_s$ part of any quantity 
$\Psi^\alpha ~ (\Phi_\alpha)$ one can use the matrices 
$\Gamma^{\pm} ~ (\tilde\Gamma^{\pm})$. Covariant generalisation of the 
latter is given by the projectors $\tilde P_{\pm} ~ (P_{\pm})$ defined 
below.

Starting from $SO(1, 9)$ vectors $D^\mu, ~ \Lambda^\mu$ and 
antisymmetric product of $\Gamma$-matrices 
\begin{eqnarray}
\tilde\Gamma^{\mu\nu}\equiv\tilde\Gamma^\mu\Gamma^\nu-
\tilde\Gamma^\nu\Gamma^\mu=2(\tilde\Gamma^\mu\Gamma^\nu+\eta^{\mu\nu})=
-2(\eta^{\mu\nu}+\tilde\Gamma^\nu\Gamma^\mu),
\end{eqnarray}
one has
\begin{eqnarray}
D^\mu\Lambda^\nu(\tilde\Gamma^{\mu\nu})^\alpha{}_\gamma 
D^\rho\Lambda^\delta(\tilde\Gamma^{\rho\delta})^\gamma{}_\beta=
4[(D\Lambda)-D^2\Lambda^2]\delta^\alpha_\beta.
\end{eqnarray}
Let $D^2=\Lambda^2=0, ~ (D\Lambda)\ne 0$. Then the matrix 
\begin{eqnarray}
\tilde K^\alpha{}_\beta\equiv\frac{1}{2(D\Lambda)}D^\mu\Lambda^\nu
(\tilde\Gamma^{\mu\nu})^\alpha{}_\beta,
\end{eqnarray}
obeys $\tilde K^\alpha{}_\gamma\tilde K^\gamma{}_\beta=
\delta^\alpha_\beta$. It allows one to define the projectors 
$(\tilde P_++\tilde P_-=1, ~ \tilde P^2_-=\tilde P_-, ~ 
\tilde P^2_+=\tilde P_+, ~ \tilde P_+\tilde P_-=0)$ 
\begin{eqnarray}
\tilde P^\alpha_{\pm\beta}=
\frac 12(\delta^\alpha_\beta\pm\tilde K^\alpha{}_\beta). 
\end{eqnarray}
It is convenient to introduce also the ``untilded'' projectors
\begin{eqnarray}
P_{\pm\alpha}{}^\beta=
\frac 12(\delta_\alpha^\beta\pm K_\alpha{}^\beta), \cr
K_\alpha{}^\beta\equiv\frac{1}{2(D\Lambda)}D^\mu\Lambda^\nu
(\Gamma^{\mu\nu})_\alpha{}^\beta, \qquad K^2=1.
\end{eqnarray}
Their properties are as follows:
\begin{eqnarray}
(P_{\pm})_\alpha{}^\beta=(\tilde P_{\mp})^\beta{}_\alpha, \cr
(D\tilde\Gamma)(\Lambda\Gamma)=-2(D\Lambda)\tilde P_-, \qquad 
(\Lambda\tilde\Gamma)(D\Gamma)=-2(D\Lambda)\tilde P_+.
\end{eqnarray}
Commutation rules for the matrices $\tilde K, ~ K$ 
\begin{eqnarray}
\tilde K^\alpha{}_\gamma(\tilde\Gamma^\mu)^{\gamma\beta}=
\frac{2}{(D\Lambda)}[D^\mu(\Lambda\tilde\Gamma)-
\Lambda^\mu(D\tilde\Gamma)]^{\alpha\beta}+
(\tilde\Gamma^\mu)^{\alpha\gamma}K_\gamma{}^\beta, \cr
\tilde K(D\tilde\Gamma)=
-\frac{1}{(D\Lambda)}[(D\Lambda)D\tilde\Gamma-
D^2\Lambda\tilde\Gamma], \cr
\tilde K(\Lambda\tilde\Gamma)=
\frac{1}{(D\Lambda)}[(D\Lambda)\Lambda\tilde\Gamma-
\Lambda^2D\tilde\Gamma], \cr
(\Lambda\Gamma)\tilde K=-
\frac{1}{(D\Lambda)}[(D\Lambda)\Lambda\Gamma-
\Lambda^2D\Gamma], \cr
(D\Gamma)\tilde K=
\frac{1}{(D\Lambda)}[(D\Lambda)D\Gamma-
D^2\Lambda\Gamma], \cr
\end{eqnarray}
imply
\begin{eqnarray}
\tilde P_{\pm}\tilde\Gamma^\mu=\tilde\Gamma^\mu P_{\pm}\pm
\frac{1}{(D\Lambda)}D^\mu(\Lambda\tilde\Gamma)\mp
\frac{1}{(D\Lambda)}\Lambda^\mu(D\tilde\Gamma), 
\end{eqnarray}
\begin{eqnarray}
\tilde P_+(D\tilde\Gamma)=
\frac{D^2}{2(D\Lambda)}\Lambda\tilde\Gamma\approx 0, \cr 
\tilde P_+(\Lambda\tilde\Gamma)=\Lambda\tilde\Gamma-
\frac{\Lambda^2}{2(D\Lambda)}D\tilde\Gamma
\approx\Lambda\tilde\Gamma, \cr
\tilde P_-(D\tilde\Gamma)=D\tilde\Gamma-
\frac{D^2}{2(D\Lambda)}\Lambda\tilde\Gamma\approx D\tilde\Gamma, \cr
\tilde P_-(\Lambda\tilde\Gamma)=
\frac{\Lambda^2}{2(D\Lambda)}D\tilde\Gamma\approx 0, 
\end{eqnarray}
\begin{eqnarray}
(D\Gamma)\tilde P_+=D\Gamma-
\frac{D^2}{2(D\Lambda)}\Lambda\Gamma\approx D\Gamma, \cr
(\Lambda\Gamma)\tilde P_+=
\frac{\Lambda^2}{2(D\Lambda)}D\Gamma\approx 0, \cr
(D\Gamma)\tilde P_-=
\frac{D^2}{2(D\Lambda)}\Lambda\Gamma\approx 0, \cr
(\Lambda\Gamma)\tilde P_-=\Lambda\Gamma-
\frac{\Lambda^2}{2(D\Lambda)}D\Gamma
\approx\Lambda\Gamma. 
\end{eqnarray}
Properties for the untilded quantities are obtained from 
Eqs.(A.16)-(A.18) by substitution $\tilde P_{\pm}\mapsto P_{\pm}, 
~ \tilde\Gamma\leftrightarrow\Gamma$.
Eqs.(A.17), (A.18) mean that the matrix $(D\tilde\Gamma)_{\alpha\beta}$ 
belong to $\tilde P_-$ subspace for the first index and to $P_+$ 
subspace for the second index. The matrix $(\Lambda\tilde\Gamma)$ has 
an opposite properties.

These properties allows one to prove two statements formulated in Sec.2. 
1a) follows from invertibility of the matrix
$(\Lambda_\mu+D_\mu){\tilde\Gamma^\mu}$: ~
$(\Lambda_\mu+D_\mu){\tilde\Gamma^\mu}[D\Gamma\Psi+
\Lambda\Gamma\Psi]\approx  
-2(D\Lambda)\Psi$. 1b) follows from Eqs.(A.17), (A.18). 1c) follows 
from manifest form of Eq.(\ref{4}) in $SO(8)$ notations
\begin{eqnarray}
\sqrt 2D^-\Psi_a-D^i\gamma^i_{a\dot a}\bar\Psi_{\dot a}=0, 
\end{eqnarray}
\begin{eqnarray} 
\sqrt 2D^+\bar\Psi_{\dot a}-D^i\bar\gamma^i_{\dot aa}\Psi_a=0.
\end{eqnarray}
From Eq.(A.20) one has 
\begin{eqnarray}
\bar\Psi_{\dot a}=\frac{1}{\sqrt 2D^+}D^i\bar\gamma^i_{\dot aa}\Psi_a.
\end{eqnarray}
Substitution of Eq.(A.21) into Eq.(A.19) gives identity on the surface 
$\Lambda^2=D^2=0$.

The statement 2) is immediate consequence of the statement 1) 
and Eqs.(A.17), (A.18).

Instead of the weak projectors (A.12), (A.13) one can  
define the strong one, starting from the matrix 
\begin{eqnarray}
\tilde K^\alpha{}_\beta\equiv\frac{1}{2b}D^\mu\Lambda^\nu
(\tilde\Gamma^{\mu\nu})^\alpha{}_\beta, \qquad 
b\equiv\sqrt{(D\Lambda)-D^2\Lambda^2},
\end{eqnarray}
instead of Eq.(A.11).

\section*{Appendix 3. Constraint algebra}
\setcounter{equation}{22}
\def\theequation{A.\arabic{equation}}
Some useful Poisson brackets are 
\begin{eqnarray}
\{D^\mu, D^\nu\}=2T\eta^{\mu\nu}\partial_\sigma\delta, \qquad 
\{D^\mu, \Lambda^\nu\}=0, \cr 
\{\Lambda^\mu, \Lambda^\nu\}=-2T\eta^{\mu\nu}\partial_\sigma\delta, \cr 
\{L_\alpha, \Lambda^\mu\}=0, \qquad
\{L_\alpha, D^\mu\}=4iT(\Gamma^\mu\partial_1\theta)_\alpha\delta,
\end{eqnarray}
where $\delta\equiv\delta(\sigma-\sigma ')$. Non zero brackets of the 
constraints (\ref{47})-(\ref{50}) consist of Eqs.(\ref{39}), (\ref{40}) 
as well as the following one: 
\begin{eqnarray}
\{G_\alpha, p_{\eta\beta}\}=-\Lambda^\mu\Gamma^\mu_{\alpha\beta}\delta, 
\qquad \{S_\alpha, p_{\chi\beta}\}=
-\Lambda^\mu\Gamma^\mu_{\alpha\beta}\delta, \cr 
\{G_\alpha, G_\beta\}=
-T[(\eta\Gamma^\mu)_\alpha(\eta\Gamma^\mu)_\beta(\sigma)+ \cr
(\eta\Gamma^\mu)_\alpha(\eta\Gamma^\mu)_\beta(\sigma ')]
\partial_\sigma\delta+
T(\eta\Gamma^\mu\partial_1\eta)\Gamma^\mu_{\alpha\beta}\delta, \cr
\{S_\alpha, S_\beta\}=
-T[(\chi\Gamma^\mu)_\alpha(\chi\Gamma^\mu)_\beta(\sigma)+ \cr
(\chi\Gamma^\mu)_\alpha(\chi\Gamma^\mu)_\beta(\sigma ')]
\partial_\sigma\delta+
T(\chi\Gamma^\mu\partial_1\chi)\Gamma^\mu_{\alpha\beta}\delta, \cr
\{G_\alpha, S_\beta\}=
-2T[(\eta\Gamma^\mu)_\alpha(\chi\Gamma^\mu)_\beta(\sigma)
\partial_\sigma\delta- \cr
2T(\eta\Gamma^\mu)_\alpha(\partial_1\chi\Gamma^\mu)_\beta\delta.
\end{eqnarray}
Note that to check Eq.(\ref{39}) one needs to use $D=10$ $\Gamma$-matrix 
identity 
\begin{eqnarray}
\Gamma^\mu_{\alpha(\beta}\Gamma^\mu_{\gamma\sigma)}=0.
\end{eqnarray}

\end{document}